\documentstyle[aps,12pt,epsfig,amssymb]{revtex}

\begin{document}
\baselineskip=0.7cm
\newcommand{\ini}{\begin{equation}}
\newcommand{\fin}{\end{equation}}
\newcommand{\inir}{\begin{eqnarray}}
\newcommand{\finr}{\end{eqnarray}}
\newcommand{\inif}{\begin{figure}}
\newcommand{\finf}{\end{figure}}
\newcommand{\bc}{\begin{center}}
\newcommand{\ec}{\end{center}}
\def\ol{\overline}
\def\pa{\partial}
\def\ra{\rightarrow}
\def\ts{\times}
\def\df{\dotfill}
\def\bs{\backslash}
\def\dg{\dagger}

$~$

\hfill DSF-01/2001

\vspace{1 cm}

\centerline{\LARGE{Leptogenesis with SU(5)-inspired mass matrices}}

\vspace{1 cm}

\centerline{\large{D. Falcone and F. Tramontano}}

\vspace{1 cm}

\centerline{Dipartimento di Scienze Fisiche, Universit\`a di Napoli,}
\centerline{Complesso di Monte Sant'Angelo, Via Cintia, Napoli, Italy}

\centerline{{e-mail: falcone@na.infn.it; tramontano@na.infn.it}}

\vspace{1 cm}

\begin{abstract}

\noindent
In the baryogenesis via leptogenesis framework the baryonic asymmetry depends
on lepton mass matrices. In a previous paper we used SO(10)-inspired mass
matrices and we found few possibilities to obtain a sufficient level of
asymmetry. In the present paper we use SU(5)-inspired mass matrices, which
also allow to check the dependence of the baryonic asymmetry on Dirac neutrino
masses. In particular, we find that the large mixing matter solution to the
solar neutrino problem, which within SO(10) gives too small asymmetry,
can now be favoured.
 
\end{abstract}

\newpage

\section{Introduction}

\noindent
The origin of the baryonic asymmetry of the universe is one of the fundamental
questions in theoretical physics \cite{rt}.
In this paper we consider the baryogenesis-via-leptogenesis mechanism
\cite{fy,luty}, where
the out-of-equilibrium decays of heavy Majorana neutrinos
produce a leptonic asymmetry which is partially converted into a baryonic
asymmetry by electroweak sphaleron processes \cite{krs}.
A minimal framework to realize this mechanism is the standard model with
heavy right-handed neutrinos, but the same mechanism works also within
unified theories, such as the SU(5) model with heavy right-handed neutrinos,
and the SO(10) model. In any case the light left-handed Majorana neutrinos are
obtained through the seesaw mechanism \cite{ss}.

The baryogenesis via leptogenesis has been discussed by several authors
\cite{bpp,cfl,bb,gks,no}.
In a previous paper \cite{ft} we have calculated the baryonic asymmetry
assuming SO(10)-inspired quark-lepton symmetry. In fact, we have used both
the relations $M_e \sim M_d$ and $M_{\nu} \sim M_u$ for the Dirac mass
matrices.
A natural motivation to consider different lepton mass matrices, in the
context of leptogenesis, is that the slightly favoured solution to the solar
neutrino problem, namely the large mixing MSW, was not able to produce enough
baryonic asymmetry.
We notice here that the SU(5) model allows more freedom for mass
matrices in that again $M_e$ is related to $M_d$, but $M_{\nu}$ is independent
from $M_u$. Moreover, the right-handed neutrino mass matrix $M_R$ is generated
as bare mass term, and not from a Yukawa coupling term with the Higgs field
that breaks the unified or intermediate gauge symmetry as happens in SO(10).
In particular, we will check the effect of several mass
hierarchies for Dirac neutrino masses.
 
In the next section we collect the relevant formulas of the
baryogenesis-via-leptogenesis mechanism. These depend on both the Dirac and the
heavy Majorana neutrino mass matrices. In the seesaw mechanism such two matrices
are related through the effective (light) Majorana neutrino mass matrix $M_L$.
Therefore, in section III we give an outline on how to determine the effective
neutrino mass matrix from neutrino oscillation data, and in section IV
we comment about Dirac mass matrices in the SU(5) model. In section V the calculation
of the baryonic asymmetry is carried out and in section VI we give our
conclusions.

\section{Baryogenesis via leptogenesis}

\noindent
A baryonic asymmetry can be generated from a leptonic asymmetry \cite{fy}.
The baryonic asymmetry is defined as \cite{kt}
\ini
Y_B =\frac{n_B-n_{\ol{B}}}{7.04 n_{\gamma}},
\fin
where $n_{B,\ol{B},\gamma}$ are number densities.
Due to electroweak sphaleron effect,
the baryonic asymmetry $Y_B$ is related to the leptonic asymmetry $Y_L$ by
\cite{ht}
\ini
Y_B=\frac{a}{a-1} Y_L
\fin
where
$$
a=\frac{8N_f+4N_H}{22N_f+13 N_H}.
$$
$N_f$ is the number of families and $N_H$ the number of Higgs
doublets. For $N_f=3$ and $N_H=1$ or $N_H=2$, it is $a \simeq 1/3$, so that
$Y_B \simeq -Y_L/2$.

The leptonic asymmetry can be written as \cite{luty}
\ini
Y_L=d ~\frac{\epsilon_1}{g^*}
\fin
where $\epsilon_1$ is a CP-violating asymmetry associated with the decay of
the lightest heavy neutrino, $d$ is a dilution factor, and $g^*=106.75$ in
the standard case or 228.75 in the supersymmetric case.
For the standard case the asymmetry $\epsilon_1$ is given by \cite{crv,bp}
\ini
\epsilon_1=\frac{1}{8 \pi v^2 (M_D^{\dg} M_D)_{11}}\sum_{j=2,3}
$Im$ [(M_D^{\dg} M_D)_{j1}]^2 f \left( \frac{M_j^2}{M_1^2} \right),
\fin
where $M_D$ is the Dirac neutrino mass matrix when $M_e$ and $M_R$ are
diagonalized, $M_i$ $(i=1,2,3)$ are the three eigenvalues of $M_R$, $v=175$
GeV is the VEV of the Higgs doublet, and
$$
f(x)=\sqrt{x} \left[1-(1+x) \ln \frac{1+x}{x}-\frac{1}{x-1} \right].
$$
In the supersymmetric version $v \ra v \sin \beta$, 
$$
f(x)=-\sqrt{x} \left[ \ln \frac{1+x}{x} +\frac{2}{x-1} \right],
$$
and a factor $4$ is included in $\epsilon_1$, due to more decay
channels. The formula (4) 
arises from the interference between the tree level and one loop
decay amplitudes of the lightest heavy neutrino, and includes vertex \cite{fy}
and self-energy \cite{ls} corrections.

The dilution factor takes into account the washout effect
produced by inverse decay and lepton number violating scattering.
A good approximation for $d$
can be inferred from refs. \cite{kt1,ap,fp}:
\ini
d=(0.1~k)^{1/2} \exp[-(4/3)(0.1~k)^{1/4}]
\fin
for $k \gtrsim 10^6$,
\ini
d=0.24/k(\ln k)^{3/5}
\fin
for $10 \lesssim k \lesssim 10^6$, and
\ini
d=1/2 k,~~d=1
\fin
for $1 \lesssim k \lesssim 10$, $0 \lesssim k \lesssim 1$, respectively,
where the parameter $k$ is
\ini
k = \frac{M_P}{1.7 v^2 32 \pi \sqrt{g^*}}\frac{(M_D^{\dg} M_D)_{11}}{M_1},
\fin
and $M_P$ is the Planck mass. In the supersymmetric case
the critical value $10^6$ for $k$ is 
lowered, but in our calculation $k$ remains always much smaller, so that only
(6) and (7) are really used.

In order to calculate the baryonic asymmetry by means of the foregoing formulas,
we have to determine $M_D$ and the heavy neutrino masses $M_i$
(and hence both $M_{\nu}$ and $M_R$). This is the aim of the next two sections.

\section{Neutrino masses and mixings}

\noindent
According to the seesaw mechanism, the light (left-handed) neutrino mass matrix
is given by the formula
\ini
M_L=-M_{\nu} M_R^{-1} M_{\nu}^T.
\fin
Of course, this formula can be inverted to give the matrix $M_R$ in terms of
a theoretical $M_{\nu}$ and a phenomenological $M_L$,
\ini
M_R=-M_{\nu}^T M_L^{-1} M_{\nu}.
\fin
In this section we discuss about $M_L$ and in the next
section about $M_{\nu}$. The effective matrix $M_L$ can be written as
\ini
M_L = U^{\dg} D_L U^*,
\fin
where $D_L= \text{diag} (m_1,m_2,m_3)$ contains the light neutrino masses
and $U$ is the lepton mixing matrix \cite{mns} in the basis with $M_e$ diagonal;
$\nu_{\alpha L}=U_{\alpha i} \nu_{i L}$ ($\alpha=e,\mu,\tau; i=1,2,3$).
The mixing matrix $U$ can be parametrizated as the CKM matrix
times a diagonal phase matrix $D=\text{diag}(e^{i \varphi_1},e^{i \varphi_2},1)$
\cite{ggab}, so that it depends on three angles and three phases.
From neutrino oscillation data we can infer the three angles \cite{pbgg} and,
assuming hierarchy $m_1 \ll m_2 \ll m_3$, also $m_2$ and $m_3$.
In particular,
for atmospheric neutrinos we use the SuperKamiokande best fit \cite{kscho}
$$
\Delta m^2_a=3.5 \ts 10^{-3} \text{eV}^2
$$
$$
\sin^2 2 \theta_a=1.0,
$$
that is maximal mixing.
For solar neutrinos we have three matter (MSW) solutions \cite{bks}, namely
the small mixing angle (SMA)
$$
\Delta m^2_s=5.4 \ts 10^{-6} \text{eV}^2
$$
$$
\sin^2 2 \theta_s=0.006,
$$
the large mixing angle (LMA)
$$
\Delta m^2_s=1.8 \ts 10^{-5} \text{eV}^2
$$
$$
\sin^2 2 \theta_s=0.76,
$$
the low-$\Delta m^2$ (LOW)
$$
\Delta m^2_s=7.9 \ts 10^{-8} \text{eV}^2
$$
$$
\sin^2 2 \theta_s=0.96,
$$
and the vacuum oscillation (VO) solution
$$
\Delta m^2_s=8.0 \ts 10^{-11} \text{eV}^2
$$
$$
\sin^2 2 \theta_s=0.75.
$$
The latest day-night and spectral data favour the LMA  solution,
but do not exclude the others \cite{ellis,flmp}.
A further information on neutrino oscillations comes from the CHOOZ experiment
\cite{chooz} which gives the bound $|U_{e3}| < 0.2$,
while $U_{e2}$ and $U_{\mu 3}$ are related to the above best fits for
atmospheric and solar neutrinos. Moreover, if light neutrino masses
are hierarchical, we have
$m_3^2 \simeq \Delta m^2_a$, $m_2^2 \simeq \Delta m^2_s$, 
$m_1 \lesssim 10^{-1} m_2$.
Therefore, we have three free phases, $\delta \equiv \arg (U_{e3})$,
$\varphi_1$, $\varphi_2$, and two bounded positive parameters,
$|U_{e3}|$, $m_1$.
Choosing values for these five parameters leads to a complete
determination of $M_L$.

\section{Mass matrices}

\noindent
In the present paper we adopt SU(5)-inspired quark and lepton mass matrices.
In the SU(5) model, without loss of generality, the matrix
$M_u$ can be taken diagonal (see for example \cite{br}).
Moreover, the charged lepton mass matrix $M_e$ is related to the down quark
matrix $M_d$. In particular, in the minimal model, where only the
Higgs multiplet {\bf 5} contributes to Dirac masses, we have $M_e = M_d^T$,
while contributions from the Higgs multiplet {\bf 45} give
$(M_e)_{ij}=-3 (M_d)_{ji}$
\cite{gj,ho}. A general approximate form for $M_d$ can be obtained from
ref.\cite{fft},
\ini
M_d=
\left( \begin{array}{ccc}
        0 & \sqrt{m_d m_s} & 0 \\
        \sqrt{m_d m_s}  & -i m_s & m_s \\
        0 & m_b/\sqrt5 & 2 m_b/\sqrt5
\end{array} \right).
\fin
It is symmetric in the 1-2 sector and leads to the famous relation
$V_{us} \simeq \sqrt{m_d/m_s} \simeq \lambda =0.22$.
Therefore, a suitable form for $M_e$ is
\ini
M_e=
\frac{m_{\tau}}{m_b} \left( \begin{array}{ccc}
        0 & \sqrt{m_d m_s} & 0 \\
        \sqrt{m_d m_s}  & 3i m_s & m_b/\sqrt5 \\
        0 & m_s & 2 m_b/\sqrt5
\end{array} \right).
\fin
The latter matrix has been obtained by taking the transpose of $M_d$, and
including a $-3$ factor in entry 2-2 in order to get good relations between
charged lepton and down quark masses \cite{gj}. 
The factor $m_{\tau}/{m_{b}}$ is an approximate factor which takes
in account the dependence of quark masses from the energy scale.
At the unification scale $m_b=m_{\tau}$, so that $M_d \sim M_e$. At a lower
scale $M_d \sim (m_b/m_{\tau}) M_e$, see ref.\cite{begn}.

It remains to consider the Dirac neutrino mass matrix $M_{\nu}$. We have
calculated the baryonic asymmetry using three hierarchies for a
diagonal form of $M_{\nu}$,
\ini
M_{\nu} =
\frac{m_{\tau}}{m_b}~ \text{diag} (\lambda ^8,\lambda ^4,1)~ m_t
\fin
\ini
M_{\nu} =
\frac{m_{\tau}}{m_b}~ \text{diag} (\lambda ^4,\lambda ^2,1)~ m_b
\fin
\ini
M_{\nu} =
\frac{m_{\tau}}{m_b}~ \text{diag} (\lambda ^2,\lambda ,1)~ m_b,
\fin
to be called 84t, 42b and 21b, respectively, where the first hierarchy is similar
to the up quark case and the second to the down quark (or charged lepton)
case. The scale of the heavy neutrino mass matrix $M_R$
is given by $M_R \sim m_t^2/m_2$ or $m_b^2/m_2$ for the SMA solution,
and by $M_R \sim m_t^2/m_1$ or $m_b^2/m_1$ for the LMA, VO, LOW solutions,
in agreement with ref.\cite{dfpl}.

\section{The baryonic asymmetry}

\noindent
As done in ref.\cite{ft}, we have random extracted the free neutrino
parameters in order to determine the
baryonic asymmetry according to the formulas included in the previous sections
(4000 points, about 2000 giving a positive asymmetry).
In figs. 1-4 we plot $Y_B$ versus $|U_{e3}|$
for the four different solar neutrino solutions
and the three hierarchies 84t, 42b, 21b.
The favoured range for the baryonic asymmetry, required for a
successful description of nucleosynthesis, is 
$Y_B=(1.7 \div 8.9) \ts 10^{-11}$ \cite{osw},
so that one can look at the region of $Y_B$ between $10^{-11}$ and $10^{-10}$.
It is clear that only the LMA solution with hierarchy 21b is fully reliable
for leptogenesis. The SMA  and LOW solutions with hierarchy 42b and the
VO solution with hierarchy 84t are acceptable for $|U_{e3}|$ tuned around
the value 0.025. The SMA solution with hierarchy 21b is acceptable for very
small values of $|U_{e3}|$.
The figures refer to the nonsupersymmetric case. However, the SU(5) model is
consistent with the phenomenology only in its supersymmetric version \cite{ua}.
In such a case the calculated
baryonic asymmetry is increased by a factor nearly 6 \cite{ft},
so that the SMA and LOW solutions with hierarchy 84t become
marginally acceptable. Also the LOW solution with hierarchy 21b becomes
acceptable.

\section{Conclusion}

\noindent
The baryonic asymmetry $Y_B$ has been calculated using a random extraction for
neutrino parameters and assuming SU(5)-inspired mass matrices.
The results depend on both the solar neutrino solution and the Dirac neutrino
mass hierarchy. The hierarchy 84t is marginally reliable for leptogenesis
for the SMA, VO and LOW solutions. The hierarchy 42b is marginally acceptable
for SMA and LOW solutions. In these cases $|U_{e3}|$ has to be tuned around
a value. The hierarchy 21b is acceptable for SMA and LOW
solutions and fully reliable for the LMA solution.

\newpage

\begin{figure}[ht]
\begin{center}
\epsfig{file=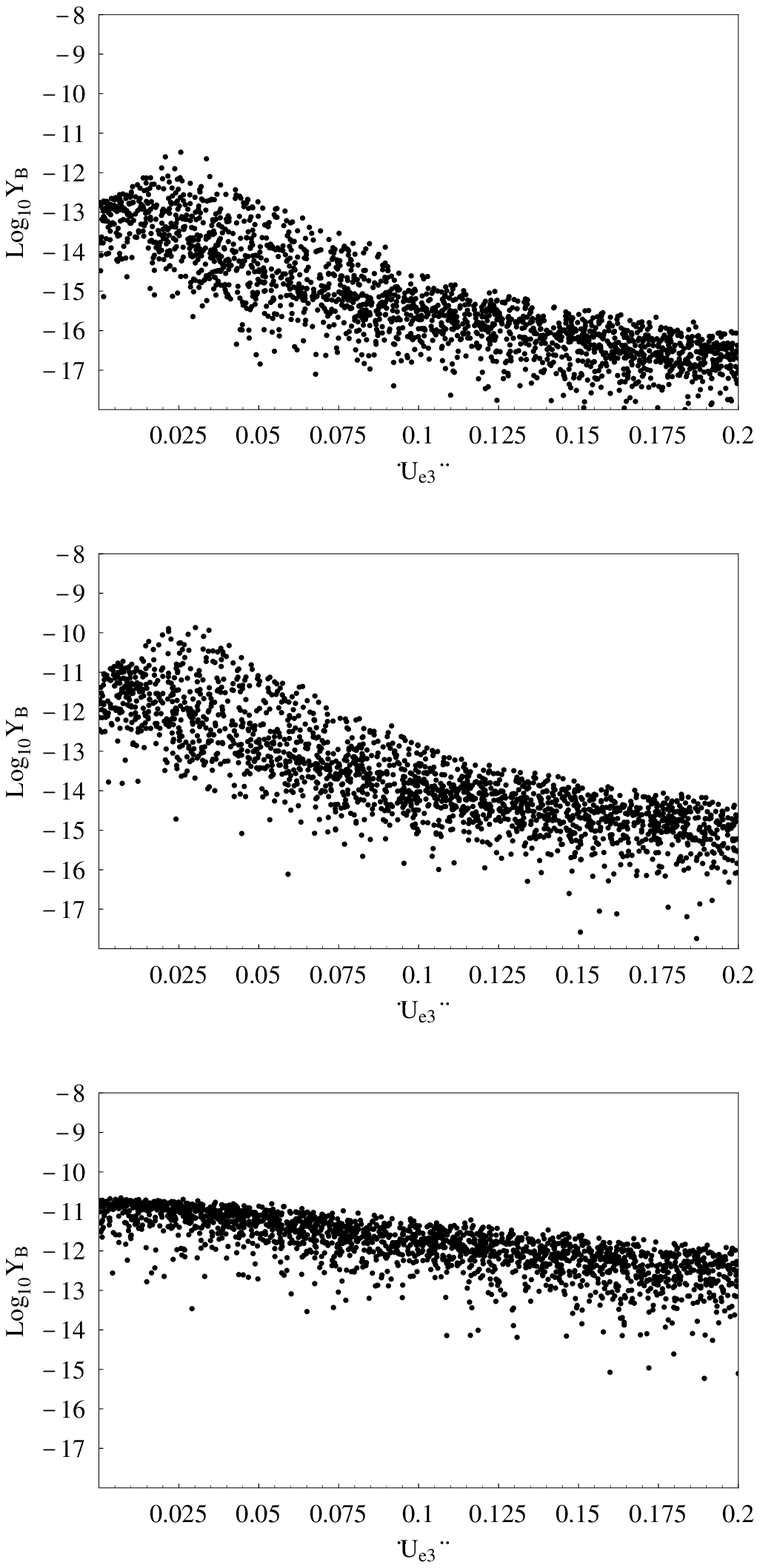,height=16cm}
\caption{The baryonic asymmetry $Y_B$ vs. $|U_{e3}|$ for SMA
and hierarchies 84t, 42b, 21b}
\end{center}
\end{figure}

\newpage

\begin{figure}[ht]
\begin{center}
\epsfig{file=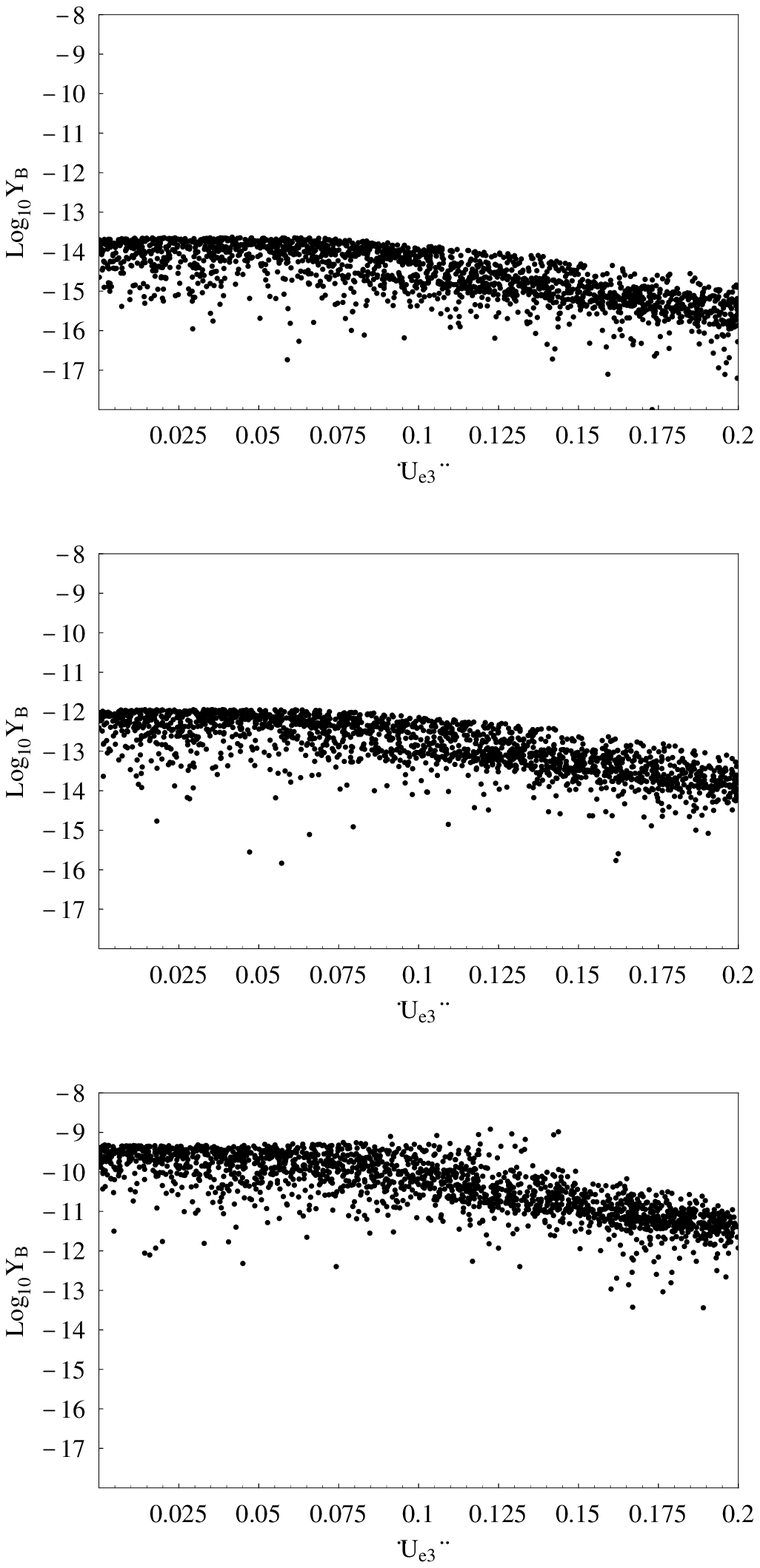,height=16cm}
\caption{The baryonic asymmetry $Y_B$ vs. $|U_{e3}|$ for LMA
and hierarchies 84t, 42b, 21b}
\end{center}
\end{figure}

\newpage

\begin{figure}[ht]
\begin{center}
\epsfig{file=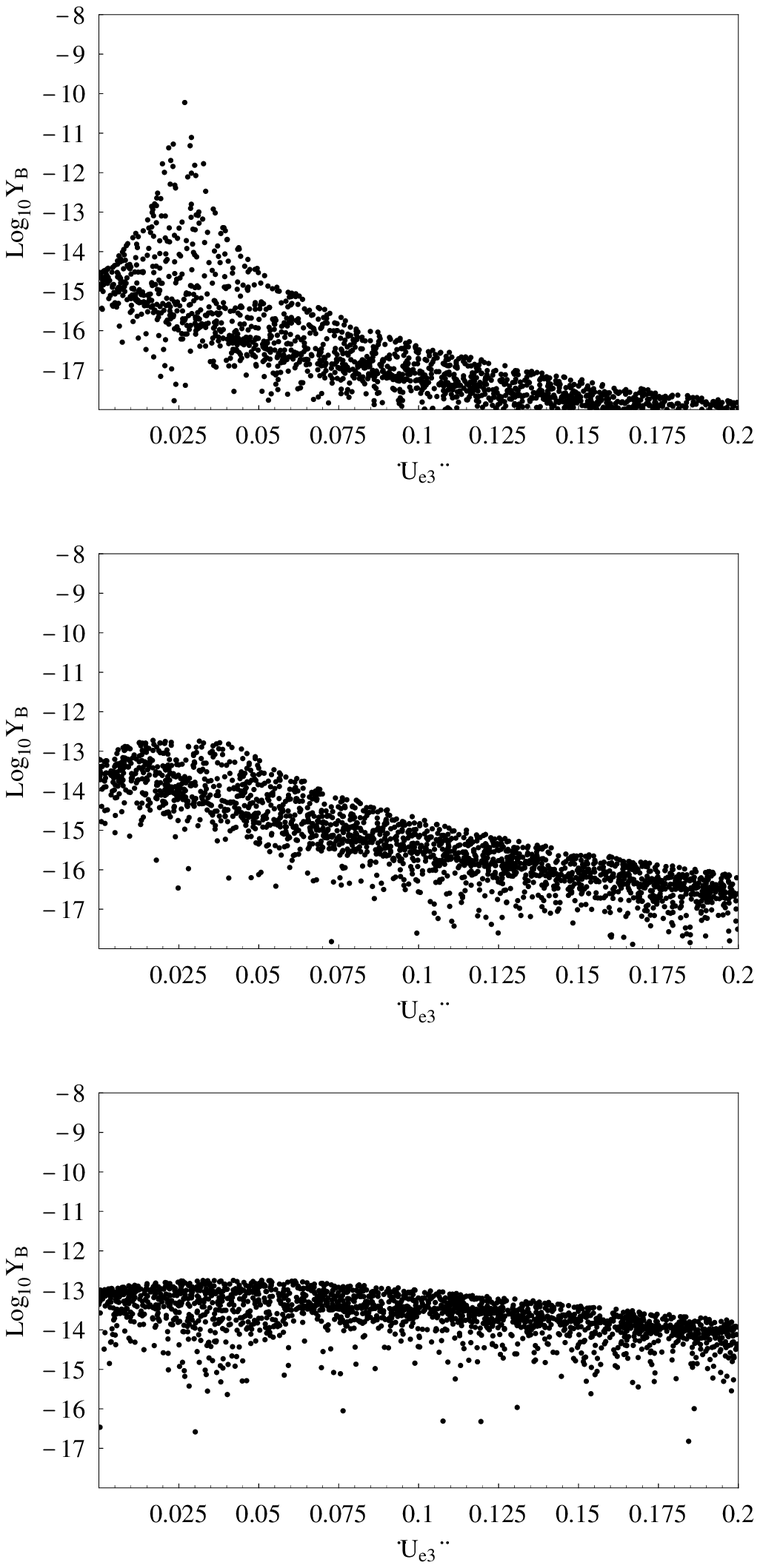,height=16cm}
\caption{The baryonic asymmetry $Y_B$ vs. $|U_{e3}|$ for VO
and hierarchies 84t, 42b, 21b}
\end{center}
\end{figure}

\newpage

\begin{figure}[ht]
\begin{center}
\epsfig{file=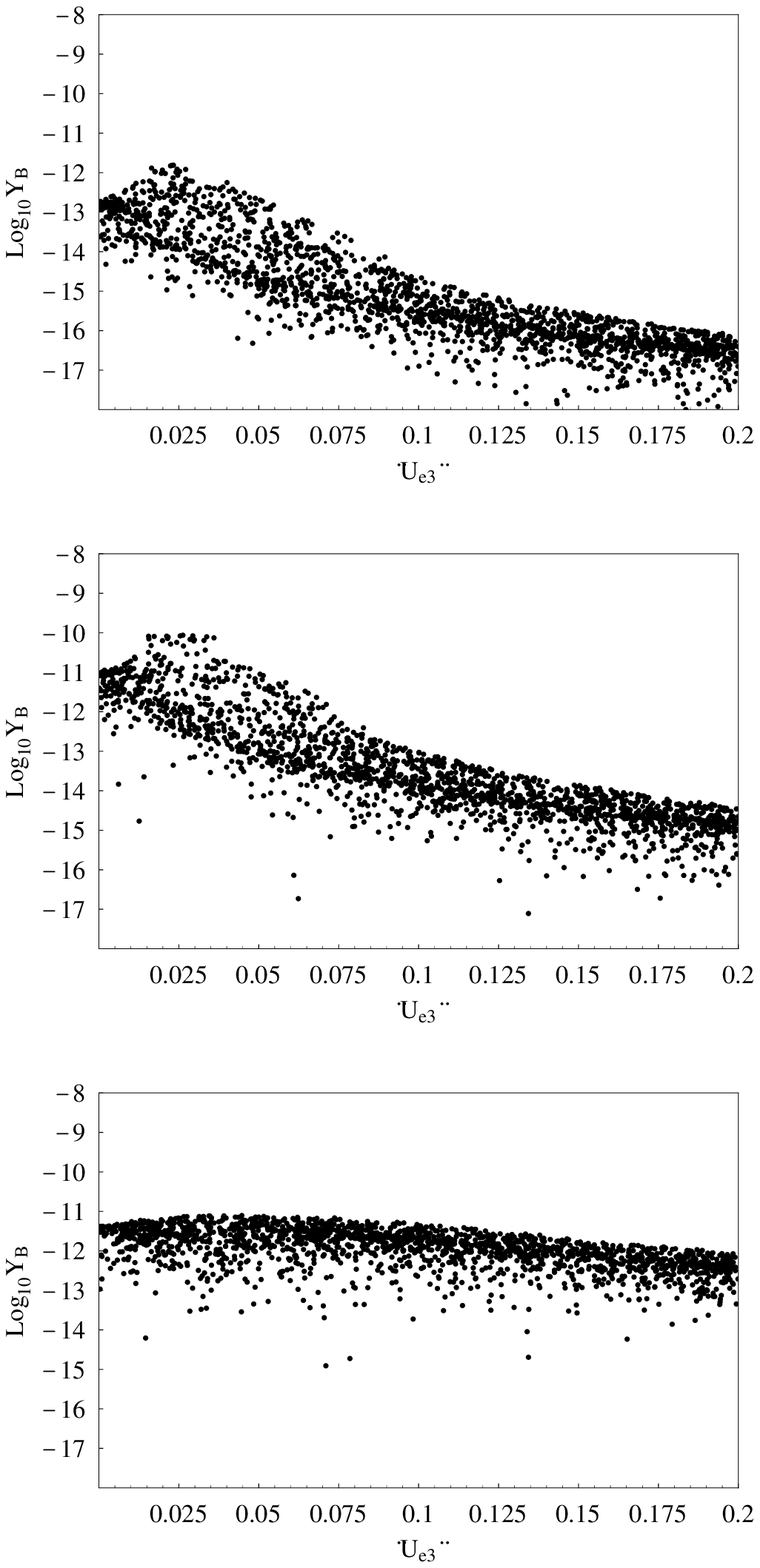,height=16cm}
\caption{The baryonic asymmetry $Y_B$ vs. $|U_{e3}|$ for LOW
and hierarchies 84t, 42b, 21b}
\end{center}
\end{figure}

\end{document}